

\hsize= 6.25in
\hoffset=.125in
\tolerance 300

\def\ll{\left\langle}
\def\rr{\right\rangle}
\def\pmb#1{\setbox0=\hbox{$#1$}%
\kern-.025em\copy0\kern-\wd0
\kern.05em\copy0\kern-\wd0
\kern-.025em\raise.0433em\box0 }

\font\twelverm = cmr12
\font\twelvei  = cmmi12
\font\twelvesy = cmsy10 at 12pt 
\font\twelvebf = cmbx12
\font\twelveit = cmti12
\font\twelvesl = cmsl12
\font\twelvett = cmtt12

\font\tenex = cmex10

\font\ninerm =  cmr9
\font\ninei  = cmmi9
\font\ninesy = cmsy9
\font\ninebf = cmbx9

\font\sixrm=cmr6
\font\sixi =cmmi6
\font\sixsy=cmsy6
\font\sixbf=cmbx6

\def\twelvept{\def\rm{\fam0\twelverm}
  \textfont0=\twelverm   \scriptfont0=\ninerm    \scriptscriptfont0=\sixrm
  \textfont1=\twelvei    \scriptfont1=\ninei     \scriptscriptfont1=\sixi
  \textfont2=\twelvesy   \scriptfont2=\ninesy    \scriptscriptfont2=\sixsy
  \textfont3=\tenex      \scriptfont3=\tenex     \scriptscriptfont3=\tenex
  \textfont\bffam=\twelvebf   \scriptfont\bffam=\ninebf
  \scriptscriptfont\bffam=\sixbf  \def\bf{\fam\bffam\twelvebf}
  \textfont\itfam=\twelveit       \def\it{\fam\itfam\twelveit}
  \textfont\slfam=\twelvesl       \def\sl{\fam\slfam\twelvesl}
  \textfont\ttfam=\twelvett       \def\tt{\fam\ttfam\twelvett}
  \normalbaselineskip=16pt plus 1pt
  \setbox\strutbox=\hbox{\vrule height11pt depth5pt width0pt
  \let\sc=\ninerm
  \normalbaselines\rm}}

\twelvept

\input epsf   

\def\footnoterule{\kern-3pt \hrule width \hsize \kern2.6pt}
\pageno=0
\footline={\ifnum\pageno>0 \hss --\folio-- \hss \else\fi}
\baselineskip 15.5pt plus 1pt

\centerline{\bf CORRELATION FUNCTIONS OF THE}
\smallskip
\centerline{\bf ONE-DIMENSIONAL RANDOM FIELD}
\smallskip
\centerline{{\bf ISING MODEL AT ZERO TEMPERATURE}\footnote{*}{This
work is supported in part by funds
provided by the U. S. Department of Energy (D.O.E.) under contract
\#DE-AC02-76ER03069.}}
\vskip 24pt
\centerline{Edward Farhi}
\vskip 12pt
\centerline{\it Center for Theoretical Physics}
\centerline{\it Laboratory for Nuclear Science}
\centerline{\it and Department of Physics}
\centerline{\it Massachusetts Institute of Technology}
\centerline{\it Cambridge, Massachusetts\ \ 02139\ \ \ U.S.A.}
\vskip 12pt
\centerline{and}
\vskip 12pt
\centerline{Sam Gutmann}
\vskip 12pt
\centerline{\it Department of Mathematics}
\centerline{\it Northeastern University}
\centerline{\it Boston, Massachusetts\ \ 02115\ \ \ U.S.A.}
\vfill
\centerline{\bf ABSTRACT}
\medskip
We consider the one-dimensional random field Ising model, where the
spin-spin coupling, $J$, is ferromagnetic and the external field is
chosen to be $+h$ with probability $p$ and $-h$ with probability
$1-p$. At zero temperature, we calculate an exact expression for the
correlation length of the quenched average of the correlation function
$\ll s_0 s_n \rr - \ll s_0\rr\ll s_n\rr$ in the case that $2J/h$ is
not an integer.  The result is a discontinuous function of $2J/h$.
When $p = {1 \over 2}$, we also place a bound on the correlation
length of the quenched average of the correlation function $\ll s_0
s_n \rr$.

\vfill
\centerline{Submitted to: {\it Physical Review B}}
\vfill
\noindent CTP \#2202  \hfill April 1993
\eject
\noindent{\bf I.\quad INTRODUCTION}
\medskip
\nobreak
The one-dimensional random field Ising model is an intriguing example
of a system with non-zero entropy at zero temperature.  The
Hamiltonian for this system is
$$H = - J \sum_i s_i s_{i+1} - \sum_i h_i s_i \eqno(1.1)$$
where the spin at each site, $s_i$, takes the values $\pm 1$ and $J>0$.
The external field $\{ h_i\}$ is frozen with the value of the field at
each site chosen as an independent random variable with, say, fifty
percent probability to be $+h$ and with fifty percent probability to
be $-h$.  For a fixed external field, $\{h_i\}$, and $h\le 2J$, there
is more than one spin configuration, $\{s_i\}$, which minimizes the
energy.  The zero temperature entropy has been calculated$^{[1,2,3]}$, and for
$2J/h$ not equal to an integer it depends only on the integer $q$,
defined by
$$q< {2J\over h}< q+1\ \ .\eqno(1.2)$$
For $2J/h$ equal to an integer, the entropy is larger than it is when
$2J/h$ is slightly less than or greater than this integer.  Thus, the
entropy is a discontinuous function of $2J/h$.  For $h>2J$,
{\it i.e.\/} $q=0$, the entropy is zero since the spins must follow
the external field.

We are interested in the correlations between the spin at site $j$ and
the spin at site $j+n$.  For a fixed external field configuration,
even at zero temperature, we must average over the different
degenerate spin configurations.  We denote this thermal average by
$\ll~~~\rr$.  An object such as $\ll s_j s_{j+n}\rr$ will depend on
the particular external field configuration, most sensitively on the
values of the external field near and between the sites $j$ and $j+n$.
What is typically measured in scattering experiments is the average
over the sample
$$G(n)\equiv \lim\limits_{N\to\infty} {1\over N} \sum\limits_j\ll s_j
s_{j+n}\rr \ \ \eqno(1.3)$$
where the sum is over the $N$ sites of the system.
By the usual ergodic arguments we can replace the spatial average by an
average over the various possible external field configurations.  This allows
us to write
$$G(n) = \overline{\ll s_0 s_n\rr} \eqno(1.4)$$
where the overbar means average over all external field configurations
generated with the probabilistic rule introduced earlier for $\{h_i\}$.

The ordinary (non-random) one-dimensional Ising model has spontaneous
magnetization at zero temperature.  Any non-zero external random
field, chosen with equal probabilities to be $\pm h$, destroys this
magnetization.  This means that for $h\not=0$,
$$\lim\limits_{N\to\infty} {1\over N} \sum\limits_j \ll s_j\rr = 0
\eqno(1.5)$$
or alternatively that $\overline{\ll s_j\rr} = 0$.  However, for a fixed
external field configuration $\ll s_j\rr$ will generally not be zero at
the site $j$. In fact, $\ll s_j\rr$ and $\ll s_{j+n}\rr$ will be correlated
so the calculation of
$$\eqalignno{
\chi(n) &= \lim\limits_{N\to\infty} {1\over N} \sum\limits_j \left[
\ll s_j s_{j+n}\rr - \ll s_j\rr \ll s_{j+n}\rr\right] &(1.6)\cr
\noalign{\hbox{which is the same as}}
\chi(n) &= \overline{\ll s_0 s_n\rr - \ll s_0\rr \ll s_n\rr} &(1.7)\cr}$$
is rather different than the calculation of $G(n)^{[4]}$.

In this paper we calculate the correlation length, $L$, of $\chi(n)$
which depends on its large $n$ behavior, i.e. $\chi(n) \sim
e^{-n/L}$. If the external random field is chosen with probability $p$
to be $h$ and with probability $1-p$ to be $-h$ we find that
$$L = {-1 \over \ln \left[ 2\left(p-p^2\right)^{1/2} \cos\left({\pi \over
q+2}\right) \right]}\eqno(1.8)$$
where $q$ is defined by (1.2) when $2J/h$ is not an integer.

We study the correlation function $G(n)$ in the case when the external
field is chosen with equal probability to be $\pm h$. We are able to
show that for any $y$ such that $y>e^{-L}$ with $L$ given by 1.8 for
$p = {1 \over 2}$, we have $\vert G(n) \vert \leq y^n$ for large enough $n$.
We argue, but do not prove, that the bound is saturated which would
imply that the two correlation functions have the same correlation
lengths.

For both $\chi(n)$ and $G(n)$ we also discuss how our calculations are
modified if $2J/h$ is an integer.

The random field Ising model is closely related to the random bond Ising model
in a uniform field. To see this, replace each $s_i$ in (1.1) with $h_is_i/h$.
The new Hamiltonian has a uniform field and random bonds $J_{i,i+1} =
Jh_ih_{i+1}/h^2$. If the $h_i$ are selected with equal probabilities to be
$\pm h$, then the bonds take the values $\pm J$ with equal probabilities and
additionally each bond is independent of the others.  However if $h_i=+h$ with
$p\not= {1\over 2}$, then the associated random bond model does not have the
bonds chosen independently. The random bond model with independently chosen
bonds is what is usually considered in the literature. For example, in
references [1,2,3] the entropy is actually calculated in a random bond model
so these results only apply to the random field model if $p={1\over 2}$.

The description$^{[2]}$ of the degenerate configurations which
contribute to the zero temperature entropy, which we give in section
II, is needed before we can attempt to calculate the correlation
functions (sections III and IV).
\goodbreak
\bigskip
\noindent{\bf II.\quad SPIN CONFIGURATIONS AT ZERO TEMPERATURE}
\medskip
\nobreak
For a given external field configuration, $\{h_i\}$, the spin configurations,
$\{s_i\}$, are those that minimize the energy (1.1).  The first term in the
energy favors agreement between adjacent sites whereas the second term prefers
the spin at a site to agree with the random field at that site.  A given
$\{h_i\}$ will not uniquely specify the $\{s_i\}$.  However, there can be
stretches of the $\{h_i\}$ that force the associated spins to take unique
values.  For example, if we find a very long stretch where all of the $h_i =
+1$, then clearly in that stretch all of the $s_i=+1$.

Consider a sequence of $k$ sites at which $h_i = -1$ and imagine that this
sequence is flanked on both sides by very long stretches where $h_i = +1$.  In
the flanking regions all $s_i=+1$.  A simple calculation shows that if
$k>2J/h$, then the spins in the sequence all match the external field {\it
i.e.\/} $s_i = -1$, whereas if $k<2J/h$ the spins are all $s_i = +1$.  Note
that if $2J/h$ is an integer and $k$ is equal to $2J/h$, then these two spin
configurations both minimize the energy.  Here we see a source of entropy
present only when $2J/h$ is an integer.

To understand how entropy arises for any non-integer value of
$2J/h>1$, consider a long stretch where $h_i = +1$, followed by a long
stretch where $h_i=-1$.  This is a deterministic situation where the
spins follow the external field.  Now at the break point imagine
inserting two additional sites where the external field takes the
values $-1$ and $+1$ so the $\{h_i\}$ configuration is $$\eqalignno{
\{h_i\} &= \cdot\ \cdot\ \cdot
\pmb{\ +\ +\ +\ +\ +\ -\ +\ -\ -\ -\ -\ -\ \cdot\ \cdot\ \cdot}\ \hbox{}\cr
\noalign{\hbox{The two spin configurations}}
\{s_i\} &= \cdot\ \cdot\ \cdot
\pmb{\ +\ +\ +\ +\ +\ +\ +\ -\ -\ -\ -\ -\ \cdot\ \cdot\ \cdot}\ \hbox{}\cr
\noalign{\hbox{and}}
\{s_i\} &= \cdot\ \cdot\ \cdot
\pmb{\ +\ +\ +\ +\ +\ -\ -\ -\ -\ -\ -\ -\ \cdot\ \cdot\ \cdot}\ \hbox{}\cr}$$
have the same energy (by symmetry) and by comparing with two other
configurations (where the spins are $-+$ and $+-$ in the middle) we see that
the two illustrated configurations minimize the energy.  This is an example
of zero-temperature entropy.

We have seen that regions of constant $h_i$ longer than $2J/h$ force the spins
to line up with the field and that there are configurations which do not
determine the spins.  We now state the general rules which dictate which
regions of the $\{h_i\}$ configuration necessarily determine the spins.

We denote a region of sites as $[\ell,r]$ if the left-most site is $\ell$
and the right-most site is $r$.  We define
$$W\left[ \ell,r\right] = {1\over h} \sum\limits^r_{i=\ell} h_i
\eqno(2.1)$$
which measures the difference between the number of sites at which the random
field is positive and the number at which it is negative in the region
$[\ell,r]$.  We further call $[\ell,r]$ an~$R^+$~region if it meets the
following three conditions:
$$R^+\ \hbox{conditions:}
\cases{
\hbox{(i)} & $W[\ell,r] >W [\ell,i] \quad \ell\le i<r$\cr\noalign{\vskip
0.2cm}
\hbox{(ii)} & $W[\ell,r] >W[i,r] \quad \ell<i\le r$\cr\noalign{\vskip 0.2cm}
\hbox{(iii)} & $W[\ell,r] \ge 2J/h$ \cr}\eqno(2.2)$$
Condition (i) says that starting from $\ell$, the number of sites at which
$h_i=+1$ minus the number at which $h_i=-1$ has a maximum in $[\ell,r]$ at $r$
and condition (iii) tells us that this maximum exceeds (or equals) $2J/h$.
Similarly, we call $[\ell,r]$ an~$R^-$~region if
$$R^-\ \hbox{conditions:}\cases{
\hbox{(i)} & $W[\ell,r]<W[\ell,i] \quad \ell\le i<r$ \cr\noalign{\vskip 0.2cm}
\hbox{(ii)} & $W[\ell,r]< W[i,r] \quad \ell<i\le r$\cr\noalign{\vskip 0.2cm}
\hbox{(iii)} & $W[\ell,r] \le - 2J/h$\ \ .\cr}\eqno(2.3)$$

Now an~$R^+$~region favors having all of the spins in $[\ell,r]$ be $+1$ over
having them all be $-1$, whereas an~$R^-$~region prefers all $-1$ spins over
all $+1$.  To guarantee that all spins in an~$R^+$~region $[\ell,r]$ be $+1$,
there should be no $R^-$ subregions of $[\ell,r]$.  We call the region
$[\ell,r]$ a $D^+$ region if it is an~$R^+$~region with no $R^-$ subregion.
Similarly, we call the region $[\ell,r]$ a $D^-$ region if it is
an~$R^-$~region with no $R^+$ subregion.

The spin at a given site will be $+1$ if the site is in a $D^+$ region.
Similarly, if a site is in a $D^-$ region the spin at that site will be $-1$.
Now a given site will be either in a $D^+$ region, a $D^-$ region, or in a
region where the spin is not forced, an $E$ region.  With our definitions a
$D^+$ region can be contained in a larger $D^+$ region.  We call a $D^+$ region
$D^+_m$ (for maximal) if it is not contained in any other $D^+$ region.
Similarly, a $D^-_m$ region is a $D^-$ region which is not a subset of a
larger $D^-$ region.  Every lattice site is either in a $D^+_m$, $D^-_m$ or
$E$ region. We cannot have consecutive $D^+_m$ regions since together they
would form a $D^+$ region which contained them both.  Similarly, two
consecutive $E$ regions will be considered as one $E$ region.

To understand what an $E$ region looks like, consider three consecutive
regions $D^+_m$, $E$ and $D^-_m$ and let $\ell$ be the left-most site of $E$
and $r$ be the right-most site {\it i.e.\/} $E=[\ell,r]$.  Consider
$W[\ell-1,i]$ as a function of $i$. Then $W[\ell-1,\ell-1] =+1$ since $\ell-1$
is the right-most site of $D^+_m$ which must end in a $+1$ site.  For any
$k\in E$, $W[\ell-1,k]\le 1$ because if this were not the case then the
$D^+_m$ region could be extended.  Similarly $W[r+1,r+1]=-1$ and
$W[k,r+1]\ge-1$ for $k\in E$.  These two inequalities imply that $W[\ell,r]=0$
which means that the entropy region has the same number of $h_i=+1$ sites as
$h_i=-1$ sites.  The function $W[\ell-1,i]$ is equal to $1$ at $i=\ell-1$ and
at $i=r$.  It can achieve the value $1$ but not exceed it at other sites in
$E$, and it also can never go below $1-2J/h$.  If it did subregions of the $E$
region would meet the conditions for being $D^+$ or $D^-$ regions.  These
properties of $W[\ell-1,i]$ will be used when we calculate the correlation
functions.

We now describe the degenerate spin configurations associated with a $D^+_m
ED^-_m$ region.  The spins are all $+1$ in $D^+_m$ and continue to be $+1$
until some point in $E$ where they switch to $-1$ and remain $-1$ through
$D^-_m$.  The last site at which $s_i$ takes the value $+1$ must be at
$i=\ell-1$ or $r$ or any other possible site in $E$ at which $W[\ell-1,i]$
happens to be $+1$.  We will illustrate this with an example momentarily.
First note that a mirror construction is used for a $D^-_m ED^+_m$ region.  It
is also possible to show that no $D^-_mED^-_m$ or $D^+_mED^+_m$ regions can
exist (when $2J/h$ is not an integer).

As an illustration, suppose $2<2J/h<3$ and we have the $\{h_i\}$ configuration
$$\hbox{\vbox{
\halign{
\hfil$#$\hfil\tabskip .05in&
\hfil$#$\hfil\tabskip .05in&
\hfil$#$\hfil\tabskip .05in&
\hfil$#$\hfil\tabskip .05in&
\hfil$#$\hfil\tabskip .05in&
\hfil$#$\hfil\tabskip .05in&
\hfil$#$\hfil\tabskip .05in&
\hfil$#$\hfil\tabskip .05in&
\hfil$#$\hfil\tabskip .05in&
\hfil$#$\hfil\tabskip .05in&
\hfil$#$\hfil\tabskip .05in&
\hfil$#$\hfil\tabskip .05in&
\hfil$#$\hfil\tabskip .05in&
\hfil$#$\hfil\tabskip .05in&
\hfil$#$\hfil\tabskip .05in&
\hfil$#$\hfil\tabskip .05in&
\hfil$#$\hfil\tabskip .05in&
\hfil$#$\hfil\tabskip .05in&
\hfil$#$\hfil\tabskip .05in&
\hfil$#$\hfil\tabskip .05in&
\hfil$#$\hfil\tabskip .05in&
\hfil$#$\hfil\tabskip 0.0in\cr
\{h_i\} =& \cdot & \cdot & \cdot
& \pmb{+} & \pmb{+} & \pmb{+} & \pmb{+} & \pmb{-} & \pmb{-} & \pmb{+}
& \pmb{+} & \pmb{-} & \pmb{+} & \pmb{-} & \pmb{-} & \pmb{+} & \pmb{-}
& \pmb{-} & \cdot & \cdot & \cdot \cr
\noalign{\vskip 0.1cm}
& \cdot & \cdot & \cdot
& 1 & 2 & 3 & 4 & 5 & 6 & 7 & 8 & 9 & 10 & 11 & 12 & 13 & 14 & 15 &
\cdot & \cdot & \cdot \cr}}}$$
where the numbers below are the site labels.
It is useful to plot $W[1,i]$ from which we can infer the values of $W$
on subregions.
\vskip 0.1cm
We can see that $[1,4]$ is a $D^+$ region whereas
$[11,15]$ which has $W[11,15]=-3$ is a $D^-$ region.  The region $[5,10]$
is an entropy region and if we look at $W[4,i]$ for
$4 \leq i \leq 10$ we see that
it is equal to 1 at $i=4$, 8 and 10.  The three degenerate spin configurations
are
$$\hbox{\vbox{
\halign{
\hfil$#$\hfil\tabskip .05in&
\hfil$#$\hfil\tabskip .05in&
\hfil$#$\hfil\tabskip .05in&
\hfil$#$\hfil\tabskip .05in&
\hfil$#$\hfil\tabskip .05in&
\hfil$#$\hfil\tabskip .05in&
\hfil$#$\hfil\tabskip .05in&
\hfil$#$\hfil\tabskip .05in&
\hfil$#$\hfil\tabskip .05in&
\hfil$#$\hfil\tabskip .05in&
\hfil$#$\hfil\tabskip .05in&
\hfil$#$\hfil\tabskip .05in&
\hfil$#$\hfil\tabskip .05in&
\hfil$#$\hfil\tabskip .05in&
\hfil$#$\hfil\tabskip .05in&
\hfil$#$\hfil\tabskip .05in&
\hfil$#$\hfil\tabskip .05in&
\hfil$#$\hfil\tabskip .05in&
\hfil$#$\hfil\tabskip .05in&
\hfil$#$\hfil\tabskip .05in&
\hfil$#$\hfil\tabskip .05in&
\hfil$#$\hfil\tabskip 0.0in\cr
\{s_i\} =& \cdot & \cdot & \cdot
& \pmb{+} & \pmb{+} & \pmb{+} & \pmb{+} & \pmb{-} & \pmb{-} & \pmb{-}
& \pmb{-} & \pmb{-} & \pmb{-} & \pmb{-} & \pmb{-} & \pmb{-} & \pmb{-}
& \pmb{-} & \cdot & \cdot & \cdot \cr\noalign{\vskip 0.1cm}
\{s_i\} =& \cdot & \cdot & \cdot
& \pmb{+} & \pmb{+} & \pmb{+} & \pmb{+} & \pmb{+} & \pmb{+} & \pmb{+}
& \pmb{+} & \pmb{-} & \pmb{-} & \pmb{-} & \pmb{-} & \pmb{-} & \pmb{-}
& \pmb{-}& \cdot & \cdot & \cdot \cr\noalign{\vskip0.1cm}
\{s_i\} =& \cdot & \cdot & \cdot
& \pmb{+} & \pmb{+} & \pmb{+} & \pmb{+} & \pmb{+} & \pmb{+} & \pmb{+}
& \pmb{+} & \pmb{+} & \pmb{+} & \pmb{-} & \pmb{-} & \pmb{-} & \pmb{-}
& \pmb{-}& \cdot & \cdot & \cdot \cr\noalign{\vskip 0.1cm}
&\cdot & \cdot & \cdot
& 1 & 2 & 3 & 4 & 5 & 6 & 7 & 8 & 9 & 10 & 11 & 12 & 13 & 14 & 15 &
\cdot & \cdot & \cdot \cr}}}$$

It is also interesting to study the same $\{h_i\}$ configuration if
$1<2J/h<2$.  In this case the only $E$ region is $[9,10]$ and the two possible
spin configurations are
$$\hbox{\vbox{
\halign{
\hfil$#$\hfil\tabskip .05in&
\hfil$#$\hfil\tabskip .05in&
\hfil$#$\hfil\tabskip .05in&
\hfil$#$\hfil\tabskip .05in&
\hfil$#$\hfil\tabskip .05in&
\hfil$#$\hfil\tabskip .05in&
\hfil$#$\hfil\tabskip .05in&
\hfil$#$\hfil\tabskip .05in&
\hfil$#$\hfil\tabskip .05in&
\hfil$#$\hfil\tabskip .05in&
\hfil$#$\hfil\tabskip .05in&
\hfil$#$\hfil\tabskip .05in&
\hfil$#$\hfil\tabskip .05in&
\hfil$#$\hfil\tabskip .05in&
\hfil$#$\hfil\tabskip .05in&
\hfil$#$\hfil\tabskip .05in&
\hfil$#$\hfil\tabskip .05in&
\hfil$#$\hfil\tabskip .05in&
\hfil$#$\hfil\tabskip .05in&
\hfil$#$\hfil\tabskip .05in&
\hfil$#$\hfil\tabskip .05in&
\hfil$#$\hfil\tabskip 0.0in\cr
\{s_i\}= & \cdot & \cdot & \cdot
& \pmb{+} & \pmb{+} & \pmb{+} & \pmb{+} & \pmb{-} & \pmb{-} & \pmb{+}
& \pmb{+} & \pmb{-} & \pmb{-} & \pmb{-} & \pmb{-} & \pmb{-} & \pmb{-}
& \pmb{-}& \cdot & \cdot & \cdot \cr\noalign{\vskip0.1cm}
\{s_i\}= & \cdot & \cdot & \cdot
& \pmb{+} & \pmb{+} & \pmb{+} & \pmb{+} & \pmb{-} & \pmb{-} & \pmb{+}
& \pmb{+} & \pmb{+} & \pmb{+} & \pmb{-} & \pmb{-} & \pmb{-} & \pmb{-}
& \pmb{-} & \cdot & \cdot & \cdot \cr\noalign{\vskip 0.1cm}
&\cdot & \cdot & \cdot
& 1 & 2 & 3 & 4 & 5 & 6 & 7 & 8 & 9 & 10 & 11 & 12 & 13 & 14 & 15 &
\cdot & \cdot & \cdot  \cr}}}$$
which do not coincide with any of the three possibilities for
$2<2J/h<3$.  If $2J/h=2$ then all five configurations are degenerate
and the entropy is larger than it is on either side of $2J/h=2$.  In
general when $2J/h$ is an integer there are even more degenerate
configurations than those one would discover by looking at $2J/h$ just
above and just below its integer value. This is because there can be
degenerate configurations which within a single $E$ region look in part like
those for $2J/h$ just above its integer value and in part look like
those for $2J/h$ just below.
\goodbreak
\bigskip
\noindent{\bf III.\quad THE CORRELATION LENGTH OF $\pmb{\chi(n)}$}
\medskip
\nobreak
Recall that
$$\chi(n) = \overline{\ll s_0 s_n\rr - \ll s_0\rr \ll s_n\rr} \eqno(3.1)$$
where $\ll~~~\rr$ is the average over different degenerate spin configurations
for fixed $\{h_i\}$ and $\overline{\phantom{\ll~~\rr}}$
is the average over $\{h_i\}$.  In this
section we determine the dominant large $n$ behavior of $\chi(n)$ for $q<2J/h
< q+1$ with $q$ an integer.  Note that for a given $\{h_i\}$, $s_0$ and $s_n$
are either determined by the external field or they are not.  We can think of
$s_0$ and $s_n$ as random variables and $\ll s_0s_n\rr - \ll s_0\rr \ll
s_n\rr$ as their statistical covariance.  If either $s_0$ or $s_n$ is forced
by the $\{h_i\}$ to take a particular value then $\ll s_0s_n\rr - \ll s_0\rr\ll
s_n\rr$ vanishes.  Thus for $\ll s_0 s_n\rr - \ll s_n\rr \ll s_0\rr$ to be
non-vanishing, both $s_0$ and $s_n$ must be in $E$ regions.  However, $\ll
s_0s_n\rr-\ll s_0\rr \ll s_n\rr$ also vanishes if $s_0$ and $s_n$ are
independent.  Now if $s_0$ and $s_n$ are in different $E$ regions, that is,
$E$ regions separated by at least one $D^+_m$ or $D^-_m$ region, then the
value of $s_0$ is independent of the value of $s_n$ and the correlation
vanishes.  For $\ll s_0 s_n\rr - \ll s_0\rr \ll s_n\rr$ to be non-zero for a
given $\{h_i\}$, both $s_0$ and $s_n$ must be in the same $E$ region.

For a  given value of $n$ we will calculate the probability, {\it i.e.\/} the
fraction of configurations $\{h_i\}$, which have 0 and $n$ in the same $E$
region.  The distribution of $\{h_i\}$ configurations is given by
assuming for simplicity that
at site $i$, $h_i = +1$ or $h_i = -1$ each with probability
one-half.  (The calculation is easily carried through with ${1\over 2}$
replaced by $p$.)
As a first step we
calculate the fraction of $E$ regions which have length $R$.
Consider an $E$ region which, for
example, begins at site 1 and ends at site $R$ and has a $D^+_m$ region to the
left and a $D^-_m$ region to the right. (Note that the sites $1$ and $R$ have
nothing to do with the sites $0$ and $n$ mentioned above.)
The function $W[0,i]$ as
discussed in the previous section has the following properties
$$W[0,0]=1\ \ ;\qquad W[0,1]=0\ \ ;\qquad W[0,R]=1\ \ ;\qquad
W[0,i]\le 1\quad\hbox{for}\quad i\in [1,R] \leqno\hbox{(i)}$$
and since $W[0,i] > 1 - 2J/h$ with $q<2J/h<q+1$ we also have
$$W[0,i]>-q\quad\hbox{for}\quad i\in [0,R]\ . \leqno\hbox{(ii)}$$
The sites just to the right of $R$ form the beginning of a $D^-_m$ region.
Therefore, the function $W[0,i]$ for $i>R$ must take the value $-q$
before it takes the value $+1$ or else the $E$ region could have been extended
beyond $R$.  So
$$W[0,i]\quad\hbox{for $i>R$ goes through $-q$ before it goes through
$+1$}\ . \leqno\hbox{(iii)}$$

The randomly generated field $\{h_i\}$ can be thought of as
determining (or as equivalent to) a random walk, RW, where position at
time $i$ changes by $h_i/h$.  The function $W[0,i]$ with $W[0,0]=1$ is
the position of the random walk at time $i$ given that at $i=0$ the
walk is at $+1$.  If we call $f_R$ the normalized probability that an
$E$ region has length $R$ we see from (i), (ii) and (iii) above that
$$\eqalign{f_R=N&\times \hbox{Prob (RW goes from 1 to 1 in $R$ steps without
hitting $+2$ or $-q$)} \cr
&\times\hbox{Prob (RW starting at 1 goes to $-q$ before returning to 1)}\cr}
\eqno(3.2)$$
where $N$ is the normalization factor.  Actually, we will calculate
the transform $$\overline{f}(\lambda)= \sum^\infty_{R=0} f_R
\lambda^R\eqno(3.3)$$ which is more useful for our purposes and from
which we can infer $f_R$.  (The $R=0$ term in the sum corresponds to
an $E$ region of zero size which occurs when there are no sites
between a $D^+_m$ region and a $D^-_m$ region.) Note that the transform
(3.3) is intimately connected to the transform of the correlation
function $\overline{\chi}(\lambda) =
\sum^\infty_{R=0}\chi(R)\lambda^R$; however we only need to calculate
(3.3) to infer the large $n$ behavior of $\chi(n)$.

Turning to the first term in (3.2), let
$$Z_1(R) = \hbox{Prob (RW goes from 1 to 1 in $R$ steps without hitting $+2$
or $-q$)}\ \ .\eqno(3.4)$$
To find this, we solve for the more general function
$$Z_j(R) = \hbox{Prob (RW goes from $j$ to $1$ in $R$ steps without hitting
$+2$ or $-q$)}\eqno(3.5)$$
and then set $j=1$.  Since a walk starts at $j$ and immediately goes to $j+1$
or $j-1$ we have for $-q+1\le j\le 1$,
$$Z_j(R) = {1\over 2} Z_{j-1} (R-1) + {1\over 2} Z_{j+1} (R-1) \ \
.\eqno(3.6)$$
By (3.5) $Z_{-q}(R)=Z_2(R)\equiv 0$ for $R\geq 0$ and $Z_1(0)=1$. If
we define $Z_2(-1)\equiv 2$ and $Z_j(-1)\equiv 0$ for $-q\leq j\leq
1$, then (3.6) holds for $R=0$ as well as for $R>0$.

We define the transform
$$\overline{Z}_j(\lambda) = \sum^\infty_{R=-1} Z_j (R)
\lambda^R\eqno(3.7)$$
which from (3.6) gives for $-q+1\le j\le 1$,
$$\overline{Z}_j (\lambda) = {1\over 2} \lambda\overline{Z}_{j-1} (\lambda) +
{1\over 2} \lambda \overline{Z}_{j+1} (\lambda) \ \ .\eqno(3.8)$$
We can solve (3.8) by making the {\it ansatz\/} that
$$\overline{Z}_j(\lambda) = \alpha u^j + \beta v^j\eqno(3.9)$$
with the boundary condition that $\overline{Z}_{-q} (\lambda) = 0$ and
$\overline{Z}_2(\lambda) = 2/\lambda$ as explained above.  The solution is
$$\overline{Z}_j (\lambda) = {2\over\lambda} \left[ {u^{q+j} - v^{q+j}\over
u^{q+2} - v^{q+2}}\right]\eqno(3.10)$$
with
$$u = {1\over\lambda} + \sqrt{{1\over \lambda^2}-1}\qquad \hbox{and}\qquad
v = {1\over\lambda} - \sqrt{{1\over\lambda^2} - 1} \eqno(3.11)$$
Thus we obtain
$$\overline{Z}_1 (\lambda)= {2\over \lambda}\left[ {u^{q+1} - v^{q+1}\over
u^{q+2} - v^{q+2}}\right] \ \ .\eqno(3.12)$$

We now return to (3.2) and we see that the second probability factor
is independent of $R$ and can therefore be absorbed in the
normalization factor $N$. Thus we have for the transform
$\overline{f}(\lambda)$ defined by (3.3),
$$\overline{f}(\lambda)=N\overline{Z}_1\,(\lambda)\eqno(3.13)$$
which by (3.12) gives
$$\overline{f}(\lambda)={2N\over\lambda}\left({u^{q+1}-v^{q+1}\over
u^{q+2}-v^{q+2}}\right)\eqno(3.14)$$
Now (3.14) can be expanded in only non-negative powers of $\lambda$ as
in (3.3).  The normalization condition $\sum\limits^\infty_{R=0}f_R=1$
is equivalent to $\overline{f}(1)=1$ which allows us to solve for $N$
and we obtain
$$\overline{f}(\lambda)= {q+2\over q+1} \  {1\over \lambda}\left(
{u^{q+1}-v^{q+1}\over u^{q+2} - v^{q+2}}\right) \ \ .\eqno(3.15)$$
Again, if we expand $\overline{f}(\lambda)$ as a power series in
$\lambda$, the coefficient $f_R$ is the probability that an
$E$ region has length $R$.

We now turn to finding
$$Q_n = \hbox{Prob(0 and $n$ are in the same $E$ region)}\ \
.\eqno(3.16)$$
The answer is
$$Q_n = N' \sum_{R>n} R f_R \left({R-n\over R}\right) \eqno(3.17)$$
where we explain each factor in turn.  The factor $N'$ contains the
probability that 0 is in an $E$ region and other $n$-independent
factors. For the $E$ region to contain 0 and $n$ it must have length
$R>n$.  The factor $Rf_R$ is proportional to the probability that an
$E$ region has length $R$ given that 0 is in it.  The factor
$\left(R-n\right)/R$ is the probability that $n$ is in an $E$ region
of length $R$ given that 0 is in it.

We can also define the transform
$$\overline{Q}(\lambda) = \sum^\infty_{n=0} Q_n\lambda^n \eqno(3.18)$$
which by (3.17) is
$$\overline{Q}(\lambda) = N' \sum^\infty_{R=1}\  \sum^{R-1}_{n=0} f_R
(R-n)\lambda^n\ \ .\eqno(3.19)$$
It is straightforward to do the sum on $n$ and then on $R$ to obtain
$$\overline{Q}(\lambda) = N' {1\over (1-\lambda)} \left[
{d\overline{f}\over d\lambda}\bigg|_{\lambda=1} - {\lambda\over
(1-\lambda)} \overline{f}(1) + {\lambda\over (1-\lambda)}
\overline{f}(\lambda)\right] \ \ .\eqno(3.20)$$
We have the explicit form of $\overline{f}(\lambda)$ through (3.15) so
the coefficients $Q_n$ in (3.18) can be determined for all $n$.  Thus
we have computed the (unnormalized) probability that 0 and $n$ are in
the same $E$ region.

The large $n$ behavior of $Q_n$ can be extracted if we know the
smallest value of $\lambda>0$, say $\lambda_*$, at which (3.20) blows
up.  This is because the expansion (3.18) will blow up first at
$\lambda=\lambda_*$ if $Q_n\sim\lambda^{-n}_*$.  Now (3.20) does not
blow up at $\lambda=1$ as can be seen by expanding
$\overline{f}(\lambda)$ about $\lambda=1$.  The only way for (3.20) to
blow up at $\lambda\not=1$ is for $\overline{f}(\lambda)$ to blow up.
{}From (3.15) we see that this can occur only if $u^{q+2} = v^{q+2}$
where again $u$ and $v$ are given by (3.11). A simple calculation
gives
$$\lambda^{-1}_*=\cos\left( {\pi \over q+2}\right)\eqno(3.21)$$
from which we conclude that, for large $n$, the probability that sites
$0$ and $n$ are in an $E$ region goes as
$$Q_n\sim\cos^n\left({\pi \over q+2}\right)\ .\eqno(3.22)$$

Given that $0$ and $n$ are in the same $E$ region we need to calculate
$\overline{\ll s_0s_n\rr - \ll s_0\rr\ll s_n\rr}^E$ where the $E$ on
the overbar denotes average only over those $\{h_i\}$ for which $0$
and $n$ are in the same $E$ region. By examining the degenerate spin
configurations in an $E$ region one can see that $\ll s_0 s_n\rr - \ll
s_0\rr \ll s_n\rr \geq 0$ for all $\{h_i\}$, so no cancellations take
place in this average. We are interested in $n$ large and the most
probable configurations contributing to this average are those for
which the $E$ region is just slightly longer than $n$ so the site $0$
and the site $n$ are near the edges of $E$. The number of degenerate
spin configurations associated with an $E$ region is proportional to
its length from which we can estimate that $\overline{\ll s_0s_n\rr -
\ll s_0\rr\ll s_n\rr}^E\sim{1\over n^2}$.

By combining (3.21) with the estimate of the previous paragraph gives,
for large $n$,
$$\chi(n)=\overline{\ll s_0s_n\rr - \ll s_0\rr\ll s_n\rr}\sim
\cos^n\left({\pi\over q+2}\right)\eqno(3.22)$$
from which we infer that the correlation length is
$$L={-1 \over\ln\cos\left({\pi\over q+2}\right)}\ .\eqno(3.23)$$

It is worth noting that if we add a constant external field, no matter
how small, then we destroy the zero temperature entropy since the
degeneracy is lifted. In this case $\chi(n)=0$ and there is also a
non-zero magnetization, i.e. $\overline{\ll s_0\rr}\not= 0$.
Alternatively we can pick the random field at each site to be $+h$
with probability $p$ and to be $-h$ with probability $1-p$, and then
$\overline{\ll s_0\rr}\not= 0$ unless $p={1\over 2}$. In this case it
is straightforward to redo the calculation of the correlation length
and we get (1.8).

If $2J/h$ is an integer, say $k$, then the calculation of the large
$n$ behavior of $\chi(n)$ changes in two ways. First, the probability
that sites $0$ and $n$ are in the same $E$ region is larger when
$2J/h$ is equal to $k$ than when $2J/h$ is slightly greater than $k$.
For example when $2J/h=1$, site $0$ and site $n$ are in the same $E$
region in the following configuration:
$$\hbox{\vbox{
\halign{
\hfil$#$\hfil\tabskip .05in&
\hfil$#$\hfil\tabskip .05in&
\hfil$#$\hfil\tabskip .05in&
\hfil$#$\hfil\tabskip .05in&
\hfil$#$\hfil\tabskip .05in&
\hfil$#$\hfil\tabskip .05in&
\hfil$#$\hfil\tabskip .05in&
\hfil$#$\hfil\tabskip .05in&
\hfil$#$\hfil\tabskip .05in&
\hfil$#$\hfil\tabskip .05in&
\hfil$#$\hfil\tabskip .05in&
\hfil$#$\hfil\tabskip .05in&
\hfil$#$\hfil\tabskip .05in&
\hfil$#$\hfil\tabskip .05in&
\hfil$#$\hfil\tabskip .05in&
\hfil$#$\hfil\tabskip .05in&
\hfil$#$\hfil\tabskip .05in&
\hfil$#$\hfil\tabskip .05in\cr
\{h_i\} = & \cdot & \cdot & \cdot
& \pmb{+} & \pmb{+} & \pmb{-} & \pmb{+} & \pmb{-} & \pmb{+} & \pmb{-}
& \pmb{+} & \pmb{-} & \pmb{+} & \pmb{+} & \cdot & \cdot & \cdot\cr
&&&&&&& 0 &&&&& n &&&&&\cr}}}$$
whereas the entire pictured region is $D^+$ if $1<2J/h<2$. But this
only approximately doubles the chance that $0$ and $n$ are in the same
$E$ region and has no effect on the correlation length.

However the typical value of $\ll s_0s_n\rr - \ll s_0\rr\ll s_n\rr$
also changes if $2J/h=k$ as opposed to $k<2J/h<k+1$. In an $E$ region
of length $n$ when $2J/h$ is not an integer there are of order $n$
configurations and $\overline{\ll s_0s_n\rr - \ll s_0\rr\ll s_n\rr}^E$
is of order $n^{-2}$. When $2J/h$ is an integer there are more
configurations. For example when $2J/h=1$ an $E$ region of length $L$
has $F_{L+2}$ configurations where $F_L$ is the $L$-th Fibonacci
number$^{[5]}$ $\left( F_1=F_2=1\ ;\ F_{L+2} = F_{L+1}+F_L\right)$. Now
for each $\{h_i\}$, $\ll s_0s_n\rr - \ll s_0\rr\ll s_n\rr$ is
still non-negative but $\overline{\ll s_0s_n\rr - \ll s_0\rr\ll
s_n\rr}^E$ is of order $F^{-2}_n$. For $1<2J/h<2$ we have that
$\chi(n) \sim \left({1\over 2}\right)^n$ as can be seen from (3.22).
For $2J/h=1$ we have $\chi(n)\sim\left({1\over 2}\right)^nF^{-2}_n$
and since $F_n\sim\left(\left(1+\sqrt5\right)/ 2\right)^n$ we infer that
$L^{-1}=\ln\left(\left(1+\sqrt5\right)^2/2\right)$.
\goodbreak
\bigskip
\noindent{\bf IV.\quad A BOUND ON THE CORRELATION FUNCTION $\pmb{G(n)}$}
\medskip
\nobreak
We are interested in the correlation length of $\overline{\ll s_0s_n\rr}$ in
the case when the external field is chosen with equal probabilities to be $\pm
h$ at each site. In this case $\overline{\ll s_0\rr}=0$. We begin by using the
symmetry of the problem to identify a class of the $\{h_i\}$ which has the
property that $\ll s_0s_n\rr$ averaged over this class is zero.  Roughly, this
is the class of $\{h_i\}$ where sites $0$ and $n$ are separated by at least 2
disjoint $D$ regions.  We then will estimate the probability that the sites
$0$ and $n$ are not in this class.  This turns out to have the same
large $n$ behavior as the probablity that sites
$0$ and $n$ are in the same $E$
region, which was relevant in calculating $\chi(n)$.

Suppose we are given a particular external field configuration
$\{h_i\}$. Let $\left[ \ell_0, r_0\right]$ be a minimal $D^+$ or $D^-$
region with $r_0 \geq 0$ and $r_0$ as small as possible. (A minimal
$D^+$ or $D^-$ region has no subregion which is a $D^+$ or $D^-$
region.) Let $\left[\ell_n, r_n\right]$ be a minimal $D^+$ or $D^-$
region with $\ell_n \leq n$ and $\ell_n$ as large as possible. Suppose
$\left[\ell_0, r_0\right]$ and $\left[ \ell_n, r_n\right]$ are disjoint,
that is $r_0<\ell_n$ (which is likely if $n$ is large). Switching all the
random field signs at $\ell_n,\ \ell_{n+1},\ \ell_{n+2},\ \ldots$ has the
effect of leaving $\ll s_0\rr$ unchanged but reversing the sign of
$\ll s_n\rr$. Note that $\ll s_0\rr\ll s_n\rr = \ll s_0s_n\rr$ in
these cases and also that performing the switch twice returns us to the
original configuration. Thus $\ll s_0s_n\rr$ averaged over all of
these configurations is zero.

Next we estimate the probability of obtaining a configuration of
external fields, $\{h_i\}$, with $r_0 \geq \ell_n$. It is these
configurations which produce a non-zero correlation function. The
reader who wishes to skip the details of this estimate should proceed
to equation 4.11 which gives the result.

We begin by calculating the distribution of $r_0$, which is the
smallest non-negative site which is the right-most site of a $D$
region. Fix $h_{-1},\ h_{-2},\ h_{-3}\ \ldots$ .  We actually
calculate the distribution of $r_0$ conditional on these values. We
will see, however, that the probability of $r_0$ being large will be
essentially the same for any choice $h_{-1},\ h_{-2},\ h_{-3}\
\ldots$ . Given the random field at the negative sites, find the
largest value of $a<0$ so that $[b,a]$ is a $D$ region for some $b<a$.
Without loss of generality assume it is a $D^+$ region. (Note that the
external field at the non-negative sites may make $[b,a]$ part of an
even larger $D$ region. However $a$ is defined only using the values
of the fields at the negative sites.)

Consider $W[a+1,r]$ as a function of $r \geq a+1$. If $W[a+1,r]$
reaches the value $1$ before it goes through $-q-1$, then the $D^+$
region $[b,a]$ can be extended. This cannot happen for $r<0$ for if it
did $a$ would not be the largest negative site ending a $D$ region on
the right. If $W[a+1,r]$ reaches the value $-q-1$ at some $r$ before
reaching the value $1$ then there is a $D^-$ region with $r$ as its
right end. Again, by assumption, this cannot happen for $r<0$. We can
see now that $r_0$ is the smallest value of $r\geq 0$ such that
$W[a+1,r]=1$ or $W[a+1,r]=-q-1$. Because $W[a+1,r]=W[a+1,-1] + W[0,r]$
for $r\geq 0$, we can say that $r_0$ is the first $r$ such that
$W[0,r]=A$ or $W[0,r]=B$ with $$\eqalignno{A&=1-W[a+1,-1]&(4.1)\cr
B&=-q-1-W[a+1,-1]\cr}$$ (where for $a=-1$ we define $W[0,-1]\equiv
0$). Thus for a fixed configuration at the negative sites, the
distribution of $r_0 \geq 0$ depends only on the single number
$W[a+1,-1]$ which obeys $-q\leq W[a+1, -1]\leq 0$.

As in the previous section we view $W[0,r]$ as equivalent to a random
walk. Let
$$Y_j(R)=\hbox{Prob(RW starting at $j$ first hits $A>0$ or $B<0$ at
step $R$)}\eqno(4.2)$$
Note that
$$\hbox{Prob}\left(r_0 = R \mid h_{-1},h_{-2},\ldots\right)=Y_0(R+1)\ .
\eqno(4.3)$$
Now
$$Y_j(R)={1\over 2}Y_{j-1}(R-1)+{1\over 2}Y_{j+1}(R-1)\eqno(4.4)$$
with the boundary conditions that $Y_A(0)=Y_B(0)=1$ and
$Y_A(R)=Y_B(R)=0$ for $R>0$. We can solve for the transform of
$Y_j(R)$,
$$\overline{Y}_j(\lambda)=\sum^\infty_{R=0} Y_j(R)\lambda^R\eqno(4.5)$$
as we did in the previous section to obtain,
$$\overline{Y}_0(\lambda)={u^A-v^A+u^{-B}-v^{-B}\over u^{q+2}-v^{q+2}}
\eqno(4.6)$$
where again $u$ and $v$ are given by (3.11) and we have used the fact
that $A-B=q+2$. Note that $\overline{Y}_0(\lambda)$ blows up for the
first time at $\lambda_*$ given by (3.21) which is independent of $A$
and $B$.

The coefficient of $\lambda^{R+1}$ in (4.6) gives the probability that
$r_0$ has the value of $R$ given a fixed $h_{-1}, h_{-2}\ldots$
which determine $A$ and $B$. Similarly we could obtain an identical
expression for the probability that $n-\ell_n$ has a given value for a
fixed $h_{n+1},h_{n+2}\ldots$ . We are interested in calculating the
probability that $r_0 \geq \ell_n$ with
$h_{-1},h_{-2}\ldots\equiv\{h_<\}$ and
$h_{n+1},h_{n+2}\ldots\equiv\{h_>\}$ both fixed. Now
$$\eqalignno{
\hbox{Prob}\left(r_0 \geq\ell_n\mid\{h_<\},\{h_>\}\right)
&=1-\hbox{Prob}\left(r_0<\ell_n\mid\{h_<\},\{h_>\}\right)\cr
&=1-\sum_{\scriptstyle i,j\geq 0\atop i+j<n}\hbox{Prob}\left(
r_0=i,n-\ell_n=j\mid\{h_<\},\{h_>\}\right)\ .&(4.7)\cr}$$
The probability that $r_0=i$ depends on the random field at sites
$\leq i$ while the probability that $n-\ell_n=j$ depends on the sites
$\geq n-j$ which do not overlap in the sum in (4.7) so the
distribution can be taken as independent. It then follows that
$$\hbox{Prob}\left(r_0\geq\ell_n\mid\{h_<\},\{h_>\}\right) =
\sum_{\scriptstyle i,j\geq 0\atop i+j\geq n}\hbox{Prob}\left(
r_0=i\mid\{h_<\}\right)\,\hbox{Prob}\left(n-\ell_n=j\mid
\{h_>\}\right)\eqno(4.8)$$
so for the purposes of our calculation we can treat the full
distributions of $r_0$ and $n-\ell_n$ as independent. Consider the
transform of the probability that $r_0+n-\ell_n$ has the value $k$:
$$\eqalignno{
\sum^\infty_{k=0}\hbox{Prob}\left(r_0+n-\ell_n=k\mid\{h_>\},
\{h_<\}\right)\lambda^k&=\left(\sum^\infty_{i=0}\hbox{Prob}\left(
r_0=i\mid\{h_<\}\right)\lambda^i\right)\times\cr
&\quad \left(\sum^\infty_{j=0}
\hbox{Prob}\left(n-\ell_n=j\mid\{h_>\}\right)\lambda^j\right)\ .
&(4.9)\cr}$$
Both transforms on the right hand side are of the form (4.6) and the
product blows up at $\lambda_*$ given by (3.21) so we can say that for
$k$ large
$$\hbox{Prob}\left(r_0+n-\ell_n=k\mid\{h_<\},\{h_>\}\right)\sim
\cos^k\left({\pi\over q+2}\right)\eqno(4.10)$$
from which we infer that
$$\hbox{Prob}\left(r_0\geq\ell_n\right)\sim\cos^n\left({\pi\over q+2}
\right)\eqno(4.11)$$
for large $n$.

The only configurations of the external field which contribute to
$\overline{\ll s_0s_n\rr}$ are those for which $r_0\geq\ell_n$. Given
a configuration with $r_0\geq\ell_n$ we expect $\ll s_0s_n\rr\sim +1\
\hbox{or}\ -1$. We have not shown that cancellations do not conspire
to make the average of $\ll s_0s_n\rr$ over those configuration with
$r_0\geq\ell_n$ of order $x^n$ with $|x|<1$. Hence we can only assert
that
$$\hbox{if}\quad y>\cos\left({\pi\over q+2}\right)\, ,\ \hbox{then}\
\left|\overline{\ll s_0s_n\rr}\right|\leq y^n\,,\
\hbox{for}\ n\ \hbox{large enough.}\eqno(4.12)$$

We can write $\overline{\ll s_0s_n\rr}$ as
$$\overline{\ll s_0s_n\rr}=\chi(n)+\overline{\ll s_0\rr\ll s_n\rr}\ ,
\eqno(4.13)$$
we know that $\chi(n)\geq0$, and from (3.23) we see that it decays as
$\cos^n\left({\pi\over q+2}\right)$. Therefore {\it if\/} $\overline{
\ll s_0\rr\ll s_n\rr}$ is non-negative we can conclude that
$\overline{\ll s_0s_n\rr}$ has as its correlation length, $L$, given by
(3.23). However we have not been able to {\it prove\/} that $\overline{
\ll s_0s_n\rr \geq0}$ although the following argument makes us believe
that it is. Consider setting the spin-spin coupling, $J$, equal to
zero, which gives $\overline{\ll s_0s_n\rr}=0$. For $J>0$ we expect
the feromagnetic coupling to induce a positive correlation between
$s_0$ and $s_n$, even in the quenched average. For this reason we
believe, but have not proven, that the correlation length of
$\overline{\ll s_0s_n\rr}$ is $L$ given by (3.23).

Finally we remark that if $2J/h=k$, an integer, the argument leading
to (4.11) again gives (4.12) with $q=k$.

\vskip 0.4in
\noindent{\bf ACKNOWLEDGEMENTS}
\medskip
We thank R.\ J.\ Birgeneau for tempting us with this
problem and A.\ N.\ Berker and M.\ Kardar for useful discussions.

\vskip 0.4in
\noindent{\bf REFERENCES}
\bigskip
\item{1.}M.~Puma and J.~F.~Fernandez, {\it Phys.~Rev.\/}
{\bf B 18} (1978) 1391.
\medskip
\item{2.}A.~Vilenkin, {\it Phys.~Rev.\/} {\bf B 18} (1978) 1474.
\medskip
\item{3.}
B.~Derrida, J.~Vannimenus and Y.~Pomeau,
{\it J.~Phys.~C: Solid State Phys.\/} {\bf 11} (1978) 4749.
\medskip
\item{4.}
B.~Derrida and H.~Hilhorst,
{\it J.~Phys.~C: Solid State Phys.\/} {\bf 14} (1981) L539.
\medskip
\item{5.}
J.~K.~Williams,
{\it J.~Phys.~C: Solid State Phys.\/} {\bf 14} (1981) 4095.

\bye